\documentclass{article}
\pdfoutput=1
\usepackage{amssymb,dsfont,mathrsfs,amsmath,listings}
\usepackage{url}
\usepackage{cite}
\usepackage{epsfig,multirow,fancybox,color,courier,helvet,mathptmx,graphicx}
\usepackage{ulem}
\usepackage{tikz}
\usepackage{quantikz} 
\usetikzlibrary{trees}
\usetikzlibrary{positioning}
\usetikzlibrary{arrows.meta}
\usetikzlibrary{positioning}
\usetikzlibrary{decorations.markings}
\usetikzlibrary{decorations.text}
\usetikzlibrary{matrix}
\usepackage[left=22mm,right=22mm,top=22mm,bottom=22mm]{geometry}

\usepackage{multicol}% http://ctan.org/pkg/multicols
\usepackage{float}
\usepackage{authblk}

\usepackage{authblk}
\usepackage{adjustbox}

\begin{document}
	\title{Quantum Computational Algorithms for\\Derivative Pricing and Credit Risk\\ in a Regime Switching Economy}
	\author[1]{Eric Ghysels}
	\author[1]{Jack Morgan}
	\author[2]{Hamed Mohammadbagherpoor} 
	\affil[1]{UNC Chapel Hill}
	\affil[2]{IBM Quantum Partner Ecosystem}
	
	\maketitle

\begin{abstract}
Quantum computers are not yet up to the task of providing computational advantages for practical stochastic diffusion models commonly used by financial analysts. In this paper we introduce a class of stochastic processes that are both realistic in terms of mimicking financial market risks as well as more amenable to potential quantum computational advantages. The type of models we study are based on a regime switching volatility model driven by a Markov chain with observable states. The basic model features a Geometric Brownian Motion with drift and volatility parameters determined by the finite states of a Markov chain.  We study algorithms to estimate credit risk and option pricing on a gate-based quantum computer. These models bring us closer to realistic market settings, and therefore quantum computing closer the realm of practical applications. 
	\end{abstract}
	
\begin{multicols}{2}
\section{Introduction}

Quantum computing (QC) algorithms for financial derivatives have been developed for a standard Black-Scholes (BS) environment with constant risk-free interest rates and constant instantaneous volatility - the so called Black-Scholes \cite{black1973pricing} model - see e.g.\ \cite{orus2019quantum, woerner2019quantum, chakrabarti2020threshold, stamatopoulos2020option} - among others.  For several decades now, the finance profession has turned its attention to models featuring a more realistic setting involving stochastic volatility - see \cite{ghysels19965, garcia2010econometrics} for literature reviews. We focus on a class of models featuring time-varying risk-free rates and instantaneous volatility involving a discrete state Markov process as an attempt to handle stochastic volatility in a QC algorithmic setting. More specifically, we study credit risk as well as option pricing risk management in a setting involving stochastic risk-free rates and volatility driven by a finite state Markov process.

The analysis in this paper relates to different strands of both the computer science and finance literatures. A number of papers have explored quantum computing algorithms for some extensions beyond the Geometric Brownian Motion model of Black-Scholes, often encountered in financial applications. In particular, \cite{kaneko2022quantum} consider the local volatility model of \cite{dupire1994pricing} which features time-varying volatility and captures some empirical features of option markets - in particular the so called smile. However, the local volatility model has some unappealing properties, notably a volatility process which is a linear function of time (and the price level) and is therefore unbounded. Also related is \cite{vazquez2021efficient} who 
introduce an approach to simplify state preparation, together with a circuit optimization technique, both of which can help reduce the circuit complexity for Quantum amplitude estimation (QAE) state preparation significantly and apply their approach to pricing European-style options under the Heston \cite{heston1993closed} stochastic volatility model. 
The Heston \cite{heston1993closed} model is the most popular stochastic volatility model for market practitioners because of its analytic tractability in computing the prices of European options. The model relaxes the constant volatility assumption in the BS model and assumes that the instantaneous variance follows a square root diffusion process with mean reversion. Its appeal is the existence of a closed-form formula for the characteristic function of the log-asset price. Moreover, European option prices can be computed using the Fourier inversion algorithm.

For pricing path-dependent options under the model, Monte Carlo simulations are often used. However, the standard Euler and Milstein time discretization simulation schemes for the continuous time diffusion models suffer from biases due to a multitude of reasons.  First, discretizations may yield negative values of the stochastic variance process which are typically set to zero before taking the square root. Second, the square root function violates the Lipschitz condition; therefore, the convergence properties of the discretization scheme may not be guaranteed. Third, the parameters of the diffusion model and those of its Euler or Milstein discretizations are not the same, but biases of the discretization schemes vanish as the sampling of the Euler/Milstein schemes increases in frequency. Financial econometricians usually handle this with dense high frequency sampling, i.e., sample say every 5 minutes in a 24-hour market, that means 288 steps for a single day on a classical computer. On a quantum computer we don’t really have the luxury (yet) to do that type of simulations, so the sampling remains coarse and therefore the discretization bias is severe. There have been numerous fixes to these issues to minimize discretization biases. A comprehensive review of these discretization schemes using various fixes can be found in \cite{lord2010comparison}. Numerical fixes are challenging even for classical digital computers as exact simulation schemes for the Heston model suffer from computationally expensive Bessel function evaluations. 
Finally, another issue with the Heston or related models is the latent volatility process. Typically it is assumed that volatility is observable. Heston took the same route and argued that near-term at-the-money BS implied volatilities could serve as plug-in values, but this also has issues - see \cite{garcia2010econometrics} for further discussion.  

Quantum computers are not up to the task of providing computational advantages along any of the aforementioned computational challenges. We therefore suggest in this paper to consider a class of stochastic processes more amenable to potential quantum computational advantages. The type of models we study are based on a regime switching volatility model driven by a Markov chain with observable states. The basic model features a GBM with drift and volatility parameters determined by the finite states of a Markov chain. 
A number of papers derived analytic expressions for European-style options is such a setting, see e.g.\ \cite{bollen1998valuing, mamon2005explicit, deshpande2008risk} and American-style options, see e.g.\ \cite{buffington2002american, zhang2004closed, huang2011methods}, among others.   However, for any other type of derivatives - such as path-dependent contracts, the pricing formula is calculated numerically through Monte Carlo simulation. 
Similar to other quantum computational implementations, we use amplitude estimation algorithms which provides a quadratic speedup compared to classical Monte Carlo methods. See \cite{herbert2021quantum} for the most recent advances.

Another application of interest is credit risk modeling. The Basel II Accord requires financial institutions to assess capital adequacy for credit, market and operational risks. Many institutions implement variations of the so-called asymptotic single risk factor model. The single latent factor accommodates cross-sectional dependence across assets but still maintains constant drift and volatility for individual obligors in a loan portfolio. The appeal is computational tractability. Adapting the single asset model of \cite{merton1974pricing} to a portfolio of credits, \cite{vasicek2002distribution} derived a function that transforms	unconditional default probabilities into default probabilities conditional on a single systematic risk factor. Like with option pricing, we introduce a new class of models suitable for today QC technology and at the same time more realistic in terms of applications of interest to financial industry quants.  Quantum computing algorithms for a standard Monte Carlo (MC) credit risk model have recently been developed \cite{egger2020credit}. Two state Markov chains are ubiquitous within credit risk calculations to mimic the impact that a bullish or bearish macroeconomic landscape has on a obligors \cite{Nielsen2011Essays}. We expand upon said model by introducing a two state Markov Chain to account for the impact of a bullish or bearish economy on credit risk.

The remainder of the paper is organized as follows. In Section \ref{sec:Markov} we introduce Markov chain models and their implementation on a gate-based quantum computer. In Section \ref{sec:credit_stat} we study QC algorithms for credit risk models of a static portfolio with regime switching. In Section \ref{sec:deriv} we look at quantum circuits for derivative pricing with regime switching. In section \ref{sec:credit_dynam} we create and evaluate a practical hybrid algorithm to determine credit risk of a dynamic portfolio.

\section{Markov Regime Switching Environment \label{sec:Markov}}

We focus on discrete state Markov chain processes. As is typical in financial engineering applications, we start with a continuous time setting. To describe this environment, let $(\Omega, \mathcal{F}, \mathds{P})$ be the underlying complete probability space and $\{X_t, t \geq 0 \}$  be an irreducible Markov chain taking values in $\mathcal{H}$ := $\{1, \ldots, M\}.$ 

The evolution of $X_t$ for all $j, k \in \mathcal{H},$ $j \neq k$ is given by 
\begin{equation}
\label{eq:Markov}
\mathds{P}\left(X_{t+ dt} = k| X_t = j\right) = \lambda_{jk} dt + o(dt), 
\end{equation}
where $\lambda_{jk}$ $\geq$ 0 $\forall$ j $\neq$ k and $\lambda_{jj}$ = - $\sum_{k \in \mathcal{H} \wedge j \neq k} \lambda_{jk}$ 
The generating $\mathds{Q}$-matrix of the chain is  $\Lambda^g$ := $\left[\lambda_{jk}\right].$ We think of $\mathds{Q}$ as the infinitesimal generator of the Markov (see e.g.\ \cite{norris1998markov} for details). Then, the discrete state Markov chain for any discrete time increment, say $\Delta t,$ can be written as $\mathscr{A}$ = $\exp{(\mathds{Q}\Delta t)},$ a version of the forward Kolmogorov equation, and according to the Chapman-Kolmogorov equation, the transition density over horizons $k \times \Delta t$ is $\mathscr{A}^k.$ Hence, unlike the typical stochastic volatility diffusion model with challenging discretization issues, we have in this case exact transition densities at all discrete sampling intervals.
Note the important difference with stochastic volatility diffusion models. For the Markov switching setup we work with discrete state, discrete time stochastic processes that have exact transition densities that are relatively easy to implement on gate-based quantum hardware, as we will show shortly. Contrast this with say the Heston model alluded to in the Introduction where simulations are far more challenging to implement on classicial as well as quantum hardware. In the remainder of the paper we will demonstrate that the ensuing risk models are appealing for implementation, and they are also more realistic in terms of mimicking financial market conditions.

We will typically think of two states, although the methods developed here extend to any arbitrary finite number of states.  Hence, for simplicity we will assume $M$ = 2, and for convenience one state is called the good economy $GE$ and the other the bad economy $BE$ state. We will assume that both states are observable, to avoid the additional complexities of filtering latent processes.

The first component of both the credit risk and option pricing models is the circuit that loads the probability distribution implied for all possible asset prices in the future into a quantum register such that each basis state represents a possible value and its amplitude the corresponding probability.
%It is important to be able to efficiently represent distributions of financial parameters on a quantum computer which might not have explicit analytical representations.
Since we opted for continuous time models with exact discrete time probability distributions - with the additional twist of being conditional on the state of a Markov process - for the purpose of quantum computing it will be most convenient to describe the Markov process via its transition matrix:
\begin{equation}
	\label{eq:transition}
	\mathscr{A} = \left[
	\begin{array}{cc}
		1 - p_{GB} & p_{GB} \\
		p_{BG} & 1 - p_{BG}
	\end{array}
	\right],
\end{equation}
where $1 - p_{GB}$ is the probability to stay in the good state,  as $p_{GB}$ is the probability to move to the the bad state. Likewise, we have 1 - $p_{BG}$ as the probability to stay in the bad state and $p_{BG}$ is the probability to move to the good state. Again, we can entertain a larger number of discrete Markov states but for our analysis it suffices to use $M$ = 2.

Unlike in the continuous time characterization, the elements of transition matrix $\mathscr{A}$ pertain to a particular discrete time sampling frequency. We look at a single-horizon setting and the probabilities correspond to the risk assessment horizon which could be weekly, bi-weekly, monthly, or beyond.

\begin{figure}[H]
	\centering
	\begin{tikzpicture}
		\node[scale=1](circ1){

			\begin{adjustbox}{width=0.8\columnwidth}
				\begin{quantikz}[column sep=0.1cm]
				\lstick{\ket{0}} & \gate{\theta_0} & \octrl{1} & \qw & \ctrl{1} & \qw & \qw & \qw & \qw & \qw & \qw & \qw & \qw\\
				\lstick{\ket{0}} & \qw & \gate{\theta_1^{(0)}} & \qw & \gate{\theta_1^{(1)}} & \octrl{1} & \qw & \ctrl{1} & \qw & \qw & \qw & \qw & \qw \\
				\lstick{\ket{0}} & \qw & \qw & \qw & \qw & \gate{\theta_2^{(0)}} & \qw & \gate{\theta_2^{(1)}} & \octrl{1} & \qw & \ctrl{1} & \qw & \qw \\
				\lstick{\ket{0}} & \qw & \qw & \qw & \qw & \qw & \qw & \qw & \gate{\theta_3^{(0)}} & \qw & \gate{\theta_3^{(1)}} & \qw \\
			\end{quantikz}
			\end{adjustbox}
		};
		\node [scale=1] [below=0.1cm of circ1] {
			\begin{adjustbox}{width=0.8\columnwidth}
			\begin{quantikz}[column sep=0.1cm]
				\lstick{\ket{0}} & \gate{\theta_0} & \qw & \ctrl{1} & \qw & \qw & \qw & \qw & \qw & \qw & \qw & \qw\\
				\lstick{\ket{0}} & \gate{\theta^{(0)}} & \qw & \gate{\theta^{(1)}-\theta^{(0)}} & \qw & \qw & \ctrl{1} & \qw & \qw & \qw & \qw & \qw \\
				\lstick{\ket{0}} &\gate{\theta^{(0)}} & \qw & \qw & \qw  & \qw & \gate{\theta^{(1)}-\theta^{(0)}} & \qw & \qw & \ctrl{1} & \qw & \qw \\
				\lstick{\ket{0}} & \gate{\theta^{(0)}} & \qw & \qw & \qw & \qw & \qw & \qw & \qw & \gate{\theta^{(1)}-\theta^{(0)}} & \qw \\
			\end{quantikz}
		\end{adjustbox}
	};
	\end{tikzpicture}
	\caption{Quantum Circuit Two-state Three-Period Markov Chain.}
	\label{fig:MCcirc}
\end{figure}
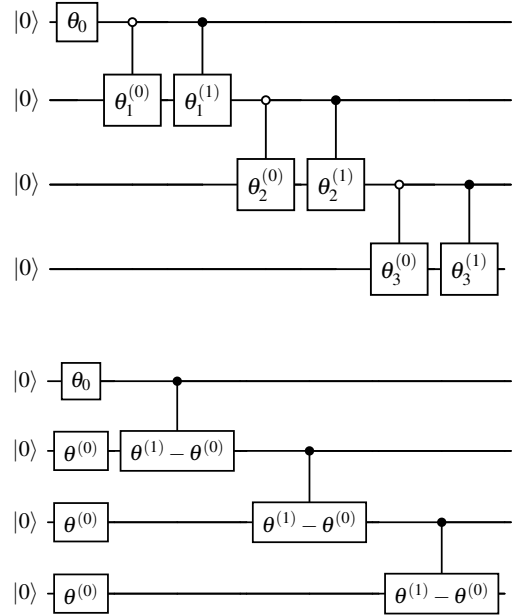

To construct the quantum Markov Chain we need the following to characterize various $R_Y(\theta)$ rotation angles. In all of our circuits we use $\ket{0}$ as the good state, and $\ket{1}$ as the bad state. First, $\theta_0$ prepares the 0th qubit in the steady state solution of $\mathscr{A}$. Next, for each qubit $i$ the $\ket{0}$ state controls a rotation of qubit $i+1$ by $\theta^{(0)}$ and the $\ket{0}$ state controls a rotation by $\theta^{(1)}$ to simulate the transition probability $p_{GB}$ and $p_{BG}$ respectively. Inspired by \cite{blank2021quantum} (their Figure 3) we have:
\begin{equation}
	\label{eq:rotations}
\begin{array}{lcl}
	\theta_0  & = & 2 \arccos (\sqrt{p_{BG}/(p_{GB} + p_{BG})}) \\
	\theta^{(0)}  & = & 2 \arccos (\sqrt{1 - p_{GB}})  \\
	\theta^{(1)}  & =  & 2 \arccos (\sqrt{p_{BG}})   
\end{array}
\end{equation}
We replace each sequence of $\theta^{(0)}$ and $\theta^{(1)}$ with an uncontrolled rotation equal to $\theta^{(0)}$, and a controlled application of $\theta^{(1)} - \theta^{(0)}$. Both circuits appear in Figure \ref{fig:MCcirc}. 

For our analysis we calculate transition probabilities based on National Bureau of Economic Research US business cycle data from 1986 to 2020.\footnote{The detailed chronology appears in \url{https://www.nber.org/research/business-cycle-dating}.} Setting $p_{GB}$ = the number of peaks divided by the number of months of expansion = 0.0097 and $p_{BG}$ = the number of troughs divided by the number of months in recession = 0.11. We assume the process to start from the steady state probability distribution. Therefore: $\theta_0$ = 1.151, 
$\theta_i^{(0)}$ = 0.311 and finally 
$\theta_i^{(1)}$ = 2.659.
%The associated {\tt Qiskit} code is in listing \ref{lst:mccode}.

%\begin{lstlisting}[language=Python, breaklines=true, caption={Example Qiskit code for parameterized circuit qc, which encodes the probabilities of every Markov Chain outcome into a register of T+1 qubits. The parameters $\theta_0$, $\theta_i^{(0)}$, and $\theta_i^{(1)}$ should be set according to equation \ref{eq:rotations}}., label=lst:mccode]
%	import numpy as np
%	from qiskit.circuit import Paramter	
%	T=3 #Set number of Time Steps
	
%	qc=QuantumCircuit(T+1) #Create circuit
	
%	'''Establish parameters'''
%	theta_naught = Parameter('theta_naught')
%	theta_0 = Parameter('theta_0')
%	theta_1 = Parameter('theta_1')
	
%	'''Prepare stable state solution'''
%	qc.ry(theta_naught, [0],'theta_naught')
	
%	'''Apply time steps'''
%	for i in range(steps):
%	qc.x([i])
%	qc.cry(theta_0, [i], [i+1], 'theta_0')
%	qc.x([i])
%	qc.cry(theta_1, [i], [i+1], 'theta_1')
%	
%	qc.measure_all()
%\end{lstlisting}

The first qubit, $q[0]$ is initialized to a superposition determined by the unconditional or steady state distribution of the Markov chain with probabilities respectively $p_{BG}/(p_{GB} + p_{BG})$ and $1 - p_{BG}/(p_{GB} + p_{BG}),$ using a $R_Y(\theta_0)$ rotation. The next qubit registers the first period two potential states based on the transition matrix characterized by controlled Y-rotations implemented with {\tt CNOT} and a single-qubit gate determined by respectively $\theta_1^{(0)}$ =  2 $\arccos (\sqrt{p_{BG}})$ and 
$\theta_1^{(1)}$ = $2 \arccos (\sqrt{1 - p_{GB}}).$ This pattern is repeated, adding a qubit for each time series observation. Since the transition matrix is time invariant, there are three parameters which drive the circuit: $p_{BG},$ $p_{GB}$ and the number of time periods $T.$ The first two determine $\theta_0,$  
$\theta_i^{(0)}$ = $\theta^{(0)}$ and  $\theta_i^{(1)}$  = $\theta^{(1)}$ $\forall$ i = 1, $\ldots, T.$ The depth of the circuit is equal to $T$ + 1 and therefore driven by the number of time periods. If the Markov chain is not homogeneous across time, then each of the $\theta_i^{(0)}$ and $\theta_i^{(1)}$ are functions of $i$.
\begin{table*}
	%	\vspace{1.0em}
	\begin{center}
		%		\footnotesize
		\begin{tabular}{ccccccc}
		%	\label{tab:MCMSE}
			Time steps & 1986-present & 1854-present & Synthetic & 1986-present & 1854-present & Synthetic \\
			& Simulator & Simulator & Simulator & Quantum & Quantum & Quantum \\
		
			3 & 2.8e-4 & 4.1e-3 & 1.3e-05 & 1.2e-4 & 2.6e-4 & 5.6e-5 \\
			6 & 5.8e-5 & 1.0e-3 & 3.8e-6 & 3.3e-4 & 5.3e-4 & 4.9e-6 \\
			12 & 9.6e-6 & 1.4e-4 & 1.9e-7 & 8.3e-4 & 2.9e-4 &  1.7e-7 \\
			24 & 2.0e-6 & 1.2e-5 & 6.2e-8 & 8.7e-5 & 5.2e-5 &  6.2e-8 \\
			
		\end{tabular}
		%--------------------------------------------------
		\caption{Mean-Squared Error (MSE) of QC Circuit Two-state $T$-Period Markov Chain evaluated on \textit{ibmq\_qasm\_simulator} simulator and \textit{ibm\_brisbane} quantum processor. The transition matrix appears in equation (\ref{eq:transition}), the circuit appears in Figure \ref{fig:MCcirc} for the case of $T$ = 3. We evaluate the accuracy of the circuit using one month transition probabilities taken from 1986-present ($p_{GB}$ = 0.0097 and $p_{BG}$ = 0.11), 1854-present ($p_{GB}$ = 0.024 and $p_{BG}$ = 0.059) as well as a synthetic case ($p_{GB}$ = 0.3 and $p_{BG}$ = 0.4). 
		\label{tab:QC_MarkovChain}}
	\end{center}
	%	\noindent{\scriptsize The entries are the Mean-Squared Errors}
\end{table*}
%To assess the performance of the circuit, we turn to Table \ref{tab:QC_MarkovChain} and Figure \ref{fig:TwoStateHistogram}. The latter displays the ideal histogram of the computational basis states for a two-state three-period Markov Chain with transition matrix appearing in equation (\ref{eq:transition}) for the case  $p_{GB}$ = .1 and $p_{BG}$ = .3 and $T$ = 3.
%\begin{figure}[H]
	%\includegraphics[width=\columnwidth]{MCoutcome.png}
	%\caption{Histogram Computational Basis States for Two-state Three-period Markov Chain. The transition matrix appears in equation (\ref{eq:transition}), the circuit appears in Figure \ref{fig:MCcirc} for the case of $T$ = 3. The histogram is computed for the case $p_{GB}$ = .1 and $p_{BG}$ = .3 and $T$ = 3.  \label{fig:TwoStateHistogram}}
%\end{figure}

To compute  the  Mean-Squared Error (MSE) for different parameter settings for the QC Circuit let us define (a) $B$ as the number of bins in the histogram of computational basis states - namely $B$ = $2^{T + 1},$ (b) $S$ the number of shots used to compute the histogram and (c) $N$ the number of Monte Carlo simulations. Moreover, let $p(b),$ $b$ = 0, $\ldots,$ $B-1$ be the true probabilities implied by the model and $o_{ns} \in $1, $\ldots,$ $B$ as the outcome of shot $s$ of iteration $n$. Then the MSE is defined as:
\begin{equation}
	\text{MSE} = \frac{1}{N} \sum_{n=1}^N \left[\frac{1}{B} \sum_{b=1}^B \left[\bar{p}_n(b) - p(b)\right]^2\right], 
\end{equation}
where $\bar{p}_n(b)$ = $ \frac{1}{S} \sum_{s=1}^S\delta_{o_{ns}, b}$ with $\delta_{o_{ns}, b}$ the Dirac delta function.

In Table \ref{tab:QC_MarkovChain} we report the  Mean-Squared Error (MSE) for different parameter settings. We consider $T$ = 3, 6, 12 and 24. In addition to the aforementioned transition probabilities, we perform the same calculation using data from 1854-present which yielded $p_{GB}$ = 0.024 and $p_{BG}$ = 0.059. We also chose sythetic probabilities $p_{GB}$ = .3 and $p_{BG}$ = .4 to demonstrate the impact of a Markov Chain with higher transition probabilities. All values were calculated N = 64 and S = 1028. Every test on quantum hardware used M3 readout error mitigation and a circuit depth optimization heuristic. Both of these protocols are standard when using \cite{runtime}. 

The results in Table \ref{tab:QC_MarkovChain} indicate that Markov chain state probabilities appear to be estimated quite accurately. In all examples the change in the MSE across multiple time steps is dominated by the exponential growth in the number of bins, and the linear decrease in number of shots expected in the most likely bins. The synthetic example has a lower MSE than the real probabilities because its probability distribution is wider, which increases the chance that two errors will cancel each other.

\section{Credit Risk with Regime Switching \label{sec:credit_stat}}
Now we know how to handle QC of Markov chain discrete state models, we turn to their first application in the context of a credit risk model. 
In a portfolio of millions of assets, it is common to group obligors together to form homogeneous groups. The most natural grouping is based on credit ratings. This can be very coarse, like pooling some of the 9 standard Moody's ratings (Aaa, Aa, A, Baa, Ba, B, Caa, Ca, and C) or more granular, like expanding say Aa into Aa1, Aa2, Aa3, etc. Another example involves tranches of structured financial products, such as Collateralized bond obligations (CBOs), Collateralized mortgage obligations (CMOs), Collateralized debt obligations (CDOs), etc. We will assume that there are $G$ homogeneous groups of obligors in the portfolio and first analyze each group separately and then cover dependence across the groups. 

We assume that the different credit risk groups $g$ share the same two-state Markov chain $X_t.$ The value of each asset belonging to group $g$ $\in$ $\{1, \ldots, G\}$  is among other things, as  discussed shortly, dependent on a discrete state Markov process describing the economic environment. 
	
%Let $\{X_t^g, t \geq 0 \}$  be an irreducible Markov chain taking values in $\mathcal{H}^g$ := $\{1, \ldots, M_g\}$ (where we set $M_g$ = 2 as noted before). The evolution of $X_t^g$ for $g$ = $1,$ $\ldots,$ $G$ is given by 
%\[
%\label{eq:Markov}
%\mathds{P}\left(X_{t+ dt}^g = k| X_t^g = j\right) = \lambda_{jk}^g dt + o(dt) \, \, \forall %j, k \in \mathcal{H}^g \, \, j \neq k 
%\]
%where $\lambda_{jk}^g$ $\geq$ 0 $\forall$ j $\neq$ k and $\lambda_{jj}^g$ = - $\sum_{k \in \mathcal{H}^g \wedge j \neq k} \lambda_{jk}^g.$ 
%The generating $\mathds{Q}$-matrix of the chain is  $\Lambda^g$ := %$\left[\lambda_{jk}^g\right]$ for $g$ = $1,$ $\ldots,$ $G.$   

Group $g$ comprises of $O_g$ obligors. For simplicity we will assume that $O_g$ = $O$ $\forall$ $g$ = $1,$ $\ldots,$ $G.$ Conditional on state $X_t$ for group $g$ the value of assets $k$ at time $t$ in logarithmic form is determined as:
	\begin{eqnarray*}
		d	\ln A^g_{k,t} (X_t) & =  & \left(\mu_g(X_t) - \frac{1}{2}\sigma_g(X_t)^2\right)dt \\
		& & \qquad \qquad \qquad + \sigma_g(X_t)d \tilde{W}^g_{k,t} \\
		d\tilde{W}^g_{k,t} & = & \sqrt{\rho_g(X_t)}dW^g_{t} + \sqrt{1 - \rho_g(X_t)} dZ^g_{k,t}
	\end{eqnarray*}
The above is a Markov switching Gaussian conditional independence model where $dW^g_{t}$ is a (group-specific) Gaussian latent factor common among all assets in group $g.$
	
Continuing with the two state setting for each group $g$ we can think of the following parameter settings:
	\begin{eqnarray*}
		\text{State } X_t & 1 \text{ (good)} & 2 \text{ (bad)} \\
		& & \\
		\mu_g(X_t)   & \text{High} & \text{Low} \\
		\sigma_g(X_t)^2  & \text{Low} & \text{High}  \\
		\rho_g(X_t)  & \text{Low} & \text{High} \\
	\end{eqnarray*}
where ``High'' and ``Low'' are calibrated according to the risk profile of group $g.$ For a portfolio of $O$ obligors the multivariate random variable $\left(\lambda_1^g, \ldots, \lambda_O^g\right)$ $\in$ $\mathbb{R}^O_+$ denotes loss given default associated to each obligor in group $g.$ Denote the probability of default for obligor $n$ in group $g$ given the latent factor $dW^g_{t}$ as $p_k(dW^g_{t}).$  We are interested in the  Value at Risk (V@R) for a given confidence level $\alpha$ $\in$ $\left(0, 1\right)$ which is defined as the smallest total loss that still has a probability greater than or equal to $\alpha$
	\begin{eqnarray*}
		\text{V@R}_\alpha(\mathcal{L}^g)  & = & \inf_{x \geq 0} \left[x \, | \, \mathbb{P}(\mathcal{L}^g \leq x) \geq \alpha \right] \\
		\mathcal{L}^g & = &  \sum_{o=1}^O \lambda_o^g p_o(dW^g_{t})
	\end{eqnarray*}
Assuming conditional independence given a single systematic risk factor, \cite{vasicek2002distribution} derived the parametric loss distribution function of an asymptotic, homogeneous credit portfolio.  We apply this asymptotic framework to each group separately, and therefore assume that each group consists of a large set of homogeneous obligors. By this we mean more specifically:
	\begin{itemize}
		\item As we already assumed, the drift, volatility and correlation $\mu_g(X_t),$ $\sigma_g(X_t),$ and $\rho_g(X_t),$ are homogeneous within each group and hence are group-specific functions of the state
		\item Obligors are assigned the same unconditional probability of default $p_g$ and the same loss given default $\lambda^g$ 
	\end{itemize}
Conditional on state $X_t$ and the latent factor the default probability is:
	\begin{equation*}
		p(dW^g_{t}, X_t) = \Phi\left(\frac{\Phi^{-1}(p_g) - \sqrt{\rho_g(X_t)}dW^g_{t}}{\sqrt{1 - \rho_g(X_t)}} \right)
	\end{equation*} 
where  $\Phi$ is the cumulative standard Gaussian density function. In the two-state case, the probability of default conditional on the latent factor and the prior state is:
	\begin{eqnarray*}
		p(dW^g_{t}|X_{t-}^g) & = &	p(X_t = 1|X_{t-}^g) 	p(X_t = 1, dW^g_{t})  \\
		& & \quad + p(X_t = 2|X_{t-}^g) 	p(X_t = 2, dW^g_{t}) 
	\end{eqnarray*}
For the subsequent analysis it will be easier to assume that the states of the Markov chain are ex post observable.  If those states are instead latent, then we need to add a (Bayesian) prior distribution $\mathcal{P}_{t-}$ on the $t_{-}$ state. In that case, the previous formula simply becomes:
		\begin{eqnarray*}
			p(dW^g_{t}|\mathcal{P}_{t-})  & = 	& \sum_j \sum_k p(X_t = j, dW^g_{t}) 	\\
			& & \quad \times p(X_t = k |X_{t-}^g = j)\mathcal{P}_{t-}(X_{t-}^g = j) 
	\end{eqnarray*}
We follow \cite{woerner2019quantum} and \cite{egger2020credit}, extending their work to the setting of Markov regime switching. Since the expected total loss $\mathbb{E}[\mathcal{L}^g]$ can be efficiently computed classically \cite{woerner2019quantum} and \cite{egger2020credit} focus on quantum algorithms to estimate $\text{V@R}_\alpha(\mathcal{L}^g).$ 

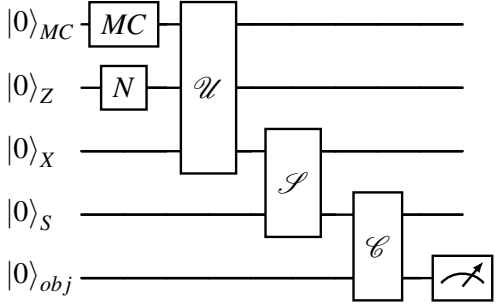
\begin{figure}[H]
	\centering
	\begin{tikzpicture}
		\node[scale=1]{
			\begin{adjustbox}{width=0.8\columnwidth}
			\begin{quantikz}[column sep={0.5cm,between origins}, row sep={0.75cm,between origins}]
				\lstick[label style={text width=0.75cm, align=left}]{$\ket{0}_{MC}$} & \gate{MC} & \qw  &\gate[3]{\mathcal{U}} & \qw & \qw & \qw & \qw & \qw & \qw \\
				\lstick[label style={text width=0.75cm, align=left}]{$\ket{0}_{Z}$} & \gate{N} & \qw & & \qw &  \qw & \qw & \qw & \qw & \qw \\
				\lstick[label style={text width=0.75cm, align=left}]{$\ket{0}_{X}$} & \qw & \qw & & \qw &  \gate[2]{\mathcal{S}} & \qw & \qw & \qw & \qw \\
				\lstick[label style={text width=0.75cm, align=left}]{$\ket{0}_{S}$} & \qw & \qw & \qw & \qw & \qw  & \qw & \gate[2]{\mathcal{C}} & \qw & \qw \\
				\lstick[label style={text width=0.75cm, align=left}]{$\ket{0}_{obj}$} & \qw & \qw & \qw & \qw &  \qw & \qw & \qw & \qw & \meter{}
			\end{quantikz}
			\end{adjustbox}
	};
	\end{tikzpicture}
	\label{fig:MCMCcirc}
	\caption{High level diagram of the $\mathcal{A}$ operator circuit. The probability of the meter measuring $\ket{1}$ is equal to the confidence interval $\alpha$ of a given loss value $\mathcal{L}$. The $MC$ gate is shown in Figure \ref{fig:MCcirc}. The $N$ gate is a normal distribution with $2^{n_Z} - 1$ values, where $n_Z$ is the number of qubits in the $Z$ register. We chose $n_Z$ = 3, mean = 3.5, and standard deviation = 1.5. $\mathcal{U}$ uses controlled y rotations to encode the likelihood of a group $g$ defaulting into the $\ket{1}$ amplitude of the corresponding qubit $X_g$ in the $X$ register. $\mathcal{S}$ computes the sum of the loss from each group in $X$ weighted by their respective value $\lambda^g$ in the $S$ register. Finally, $\mathcal{C}$ flips the objective qubit if the total loss is greater than or equal to $\mathcal{L}$. The normal distribution, weighted sum, and comparator gates are from Qiskit's circuit library \cite{gadi_aleksandrowicz}.}
\end{figure}

We first review the cases without Markov switching considered by \cite{egger2020credit}.  In their case, mapping the CDF of the total loss to a quantum operator $\mathcal{A}$ requires three steps. Each step corresponds to a quantum operator:
\begin{itemize}
	\item First, $\mathcal{U}$ loads the uncertainty model.
	\item Second, $\mathcal{S}$ computes the total loss into a quantum register with $n_S$ qubits.
	\item Last, $\mathcal{C}$ flips a target qubit if the total loss is less than or equal to a given level $x$ which is used to search for V@R.
\end{itemize}
At a high level we have $\mathcal{A}$ = $\mathcal{CSU}$ with the corresponding circuit appearing in Figure \ref{fig:MCMCcirc}, which extends Figure 1 from \cite{egger2020credit} to a Markov regime switching setting. Regarding the $\mathcal{U}$ operator, one can encode the default events for each obligor $k$ in the state of a corresponding qubit by applying to qubit $k$ a $Y$-rotation $R_Y(\theta_g)$ with angle $\theta_g$ = $2 \arcsin(	p(dW^g_{t}, X_t)).$ Note that the default probabilities and sensitivities are Markov chain state-dependent, and therefore in the case of two states we need to encode two probability schemes, leading to the following modification of the original algorithm (assuming as noted earlier that Markov states are ex post observable):
\begin{equation}
	\mathcal{U} = \sqrt{p(X_t = 1|X_{t-}^g)} \mathcal{U}_1 + \sqrt{(1 - p(X_t = 1|X_{t-}^g))} \mathcal{U}_2
\end{equation}	
where $\mathcal{U}_j$ encodes the default probabilities in Markov state $X_t$ = $j.$ In our case, $\mathcal{U}_{GE}$ is constructed using default probability $GE p_g$ and sensitivity $GE p_g$, while $\mathcal{U}_{GE}$ uses $GE p_g$ and $GE p_g$ respectively. Like \cite{egger2020credit}, we use a truncated and discretized approximation of $dW^g_k$ with $2^{n_Z} - 1$ values, considering an affine mapping $dW^g_k$ = $a_w k$ + $b_w$ from $k$ $\in$ $\{0, 1, \ldots, 2^{N_Z} - 1\}$ to the desired range of values for the latent factor. The probability of default for a given group $g$ could be encoded into the $\ket{1}$ amplitude of the group's corresponding qubit $X^g$ using controlled y rotations with $\theta_p^g(z) = 2 \arcsin\left(\sqrt{p\left(dW_t^g,X_t\right)}\right)$ controlled by each possible state in our mapping of $dW^g_k$. We classically compute the first order Taylor approximation $\theta_p^k \approx a_g z + b_g$ which can be efficienly mapped with a Pauli polynomial circuit. With the redefined $\mathcal{U}$ operator, the other operators $\mathcal{S}$ and $\mathcal{C}$ are the same as in \cite{egger2020credit}. The Markov chain operations for a single group can be implemented (assuming 2 states) using a controlled $n$th root Pauli X-gate $C\sqrt[n]{X}$ (see \cite{nikolov2019markov}). We show the results of this circuit for a variety of parameters in Table \ref{tab:CredRiskRes}. The procedure computes the probability $\alpha$ for an input loss value $\mathcal{L}$.  We expand on \cite{egger2020credit} by implementing this circuit on a real quantum processor, as opposed to a simulator, the results of which can be seen in Table \ref{tab:CredRiskRes}.

In most applications, the quantity of interest is $\mathcal{L}$ given $\alpha$. Our circuit can be used to find $\mathcal{L}$ in a simple binary search which scales in $\mathcal{O}\left(\log n\right)$ time \cite{egger2020credit}.

\begin{table*}
	\begin{center}
		\begin{tabular}{lcccc}
			Regime Switching & $GE p_g$ & $BE p_g$ & $GE \rho_g$ & $BE \rho_g$ \\
			& & & &  \\
			No  &  [0.15,0.25] & -- & [0.1, 0.05]  & -- \\
			Yes &  [0.1,0.2] & [0.15,0.25] & [0.1, 0.05]  & [0.15,0.1] \\
		\end{tabular}
		\caption{\label{tab:Params} Model parameters for the static portfolio Credit Risk calculations whose results are reported in Table \ref{tab:CredRiskRes}. We label the six month default probability $p_g$ and sensitivity $\rho_g$ for the good economy (GE) and bad economy (BE) regimes respectively. In all experiments $\lambda^g$ = [1,2]. While the first model has no regime switching, for simplicity we list the only the set of default probabilies and sensitivies in the GE column. In the regime switching model, the default probability and sensitivity for a group of obligors are larger in the bad economy regime, and our approximation of $p(dW^g_t,X_t)$ is divided by $T$ in order to determine the one month default probability.}
	\end{center}
\end{table*}	
\begin{table*}
	\begin{center}
		%	\vspace{1.0em}
		%		\footnotesize
		\begin{tabular}{lccc}
		Processor & No Switching & 1986-present & Synthetic  \\
			& & & \\
			 Simulator & 2.1e-3 & 2.6e-4 & 3.5e-3 \\
			 Quantum & 1.1e-2 & 3.7e-27 & 4.7e-2 \\
		\end{tabular}
		%--------------------------------------------------
	%	\noindent{\scriptsize The entries are the Mean-Squared Errors}
		\caption{\label{tab:CredRiskRes} The MSE of static portfolio Credit Risk calculations. We evaluate a model without regime switching, as well as two regime switching models using the 1986-present and synthetic transition probabilities. We run on both models the noiseless \textit{ibmq\_qasm\_simulator}, as well as the \textit{ibm\_kolkata} Falcon quantum processor. The portfolio parameters for both cases are listed in \ref{tab:Params}.}
	\end{center}
\end{table*}

\section{Derivative Pricing with Regime Switching \label{sec:deriv}}

Financial derivatives, such as option contracts, are valid for a pre-determined period of time and their value at the expiration date is called the payoff. Option pricing consists of determining the payoff at the expiration date in the future and then discount that value to determine  its fair value today. Here we study derivative pricing in a setting involving a state-of-the-economy process $X_t$ which is a discrete time discrete state space Markov chain. Once the state of economy is given, stock prices evolve according to some laws of motion which ultimately determine the value of an option.

More formally, let $(\Omega, \mathcal{F}, \mathds{P})$ be the underlying complete probability space and $\{X_t, t \geq 0 \}$  be an irreducible Markov chain taking values in $\mathcal{H}$ := $\{1, \ldots, M\}$ as described in equation  (\ref{eq:Markov}).

An instantaneous risk-free asset is available defined by the mapping $r:$ $\mathcal{H}$ $\rightarrow$ $[0, \infty)$ and therefore $r(X_t)$ is an irreducible Markov chain taking values in $\mathcal{R}$ := $\{r(1), \ldots, r(M)\}$ with the same generating matrix $\Lambda.$ Also of interest is the process:
\[
d B_t = r(X_t) B_t dt \quad B_0 = 1 \quad \Rightarrow B_t = \exp{(\int_0^t r(X_t)dt)}.
\]
The time variation of the risky asset price $\{S_t, t \geq 0 \}$ under $\mathds{P}$ is governed by:
\[
d S_t = \mu(X_t) S_t dt + \sigma(X_t) S_t d W_t,
\]
with $W_t$ a standard Weiner process and the functions $\mu$ and $\sigma$ are Lipschitz and satisfy:
$\int_{0}^{t} |\mu(X_t)| d t$ $<$ $\infty,$ 
$\int_{0}^{t} |\sigma(X_t)|^2 d \tau$ $<$ $\infty$  almost surely.

From \cite{ghosh1997ergodic} we know that the joint process $\{S_t,X_t\}$ is a Feller Markov process. In addition, let $\mathcal{F}_t$ := $\sigma \left[ S_t, X_t, t \geq 0 \right] $ be the natural sigma filtration assumed to be right continuous and $\mathds{P}$-complete. The sigma filtration $\mathcal{F}_t$ induces a conditional probability measure $\mathds{P}_t.$

Under complete markets there is a unique risk neutral density exists and is
related to the physical probability measure $\mathds{P}_t$ via the Radon-Nikodym derivative:
$\frac{d \mathds{P}^*_t}{d \mathds{P}_t}$ = $\exp{\left[-\int_0^t\frac{\mu(X_s) - r(X_s)}{\sigma(X_s)}dW_s - \frac{1}{2} \int_0^t\left(\frac{\mu(X_s) - r(X_s)}{\sigma(X_s)}\right)^2ds\right]}.$
Derivatives are priced according to
\[
d S_t = r(X_t) S_t dt + \sigma(X_t) S_t d W_t^*
\]
where  $W_t^*$ = $W_t$ + $\int_0^t\left(\mu(X_s) - r(X_s)\right)/\sigma(X_s)ds.$ 
To proceed we characterize the asset price dynamics in terms of log returns, which is accomplished via applying It\^o's Lemma:
\[ \label{eq:log_return}
d \ln S_t = \left(r(X_t) - \frac{1}{2}\sigma(X_t)^2\right)dt + \sigma(X_t)dW^*_t.
\]

Consider a European-style option with payoff at some future time $T$ and strike price $K.$ The future payoff of such a contract is $(S_T - K)^+$ and current market price for the option is equal to 
\[
C(t,T,K,X_t,S_t) = \mathbb{E}^*\left[e^{-\int_t^Tr(X_s)ds}\left(S_T - K\right)^+|\mathcal{F}_t \right]
\]
\cite{mamon2005explicit} and \cite{deshpande2008risk} show that the option price satisfies a system of Black–Scholes partial differential equations with weak coupling.

In the special case of two states, \cite{zhu2012new}  derive an easy exact solution for pricing European-style options 
The virtue of this model, as well as the credit risk model in the previous section is that while they are formulated in continuous time, we do know, conditional on information at time $t,$ the implied discrete time probability density of $S_T$ for any $T$ $>$ $t.$ Namely, this distribution is log-normal and is the reason why the Black-Scholes option pricing model has a known analytical solution. Here, however, we are dealing with an additional source of uncertainty, namely the discrete state Markov chain. Conditional on observing the state at time $t$ we have exact solutions for each regime. This discrete state stochastic environment facilitates implementation on quantum computers.

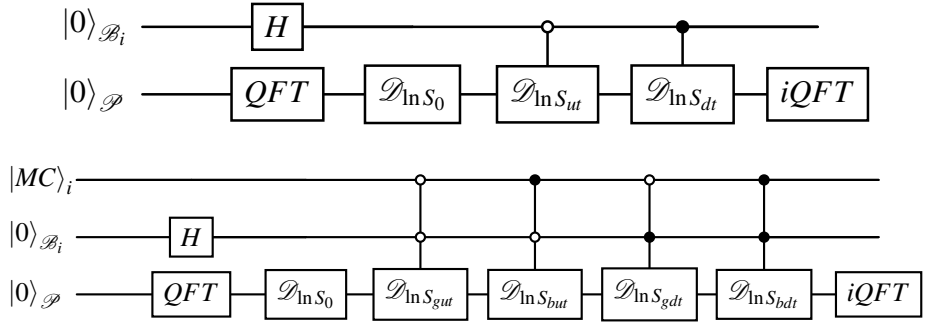
\begin{figure*}
	\centering
	\begin{tikzpicture}
		\node[scale=0.8](circ1){
			\begin{adjustbox}{width=0.8\linewidth}
			\begin{quantikz}[row sep={0.75cm,between origins}, column sep={1.5cm,between origins}]
				\lstick[label style={text width=0.75cm, align=left}]{$\ket{0}_{\mathcal{B}_i}$} & \gate{H} & \qw & \octrl{1} & \ctrl{1} & \qw \\
				\lstick[label style={text width=0.75cm, align=left}]{$\ket{0}_{\mathcal{P}}$} & \gate{QFT} & \gate{\mathcal{D}_{\ln S_0}} & \gate{\mathcal{D}_{\ln S_{ut}}} & \gate{\mathcal{D}_{\ln S_{dt}}} & \gate{iQFT}
			\end{quantikz}
			\end{adjustbox}
		};
		\node[scale=.9][below=0.1cm of circ1]{
			\begin{adjustbox}{width=0.8\linewidth}
			\begin{quantikz}[row sep={0.75cm,between origins}, column sep={1.5cm,between origins}]
				\lstick[label style={text width=0.75cm, align=left}]{$\ket{MC}_{i}$} & \qw & \qw & \octrl{1} & \ctrl{1} & \octrl{1} & \ctrl{1} & \qw \\
				\lstick[label style={text width=0.75cm, align=left}]{$\ket{0}_{\mathcal{B}_i}$} & \gate{H} & \qw & \octrl{1} & \octrl{1} & \ctrl{1} & \ctrl{1} & \qw \\
				\lstick[label style={text width=0.75cm, align=left}]{$\ket{0}_{\mathcal{P}}$} & \gate{QFT} & \gate{\mathcal{D}_{\ln S_0}} & \gate{\mathcal{D}_{\ln S_{gut}}} & \gate{\mathcal{D}_{\ln S_{but}}} & \gate{\mathcal{D}_{\ln S_{gdt}}} & \gate{\mathcal{D}_{\ln S_{bdt}}} & \gate{iQFT}
			\end{quantikz}
			\end{adjustbox}
		};
	\end{tikzpicture}
	\caption{\label{fig:ChangeCirc} Circuit diagram of one time step of $\mathcal{C}$ in the case without (top) and with (bottom) regime switching. The gates $\mathcal{D}_i$ consists of a single Y-rotation on each qubit, the angle of which is determined by the value being added and the binary place the qubit represents. The QFT and iQFT gates are a standard Fourier Transform and its inverse respectively. The values of $\ln S_{xy}$ come from equation \ref{eq:S_t}, while the values of $\ln S_{xyz}$ can be derived using \ref{eq:log_return} for x $\in$ good economy $g$ or bad ecomony $b$, and y $\in$ up $u$ or down $d$ Brownian motion.}
\end{figure*}

We compute the case without the Markov chain by using the portion of the circuit outlined in \cite{Ramos_Calderer_2021} that simulates the change in the asset price, and the standard Linear Amplitude payoff function available in \textit{Qiskit} \cite{gadi_aleksandrowicz}. In this model, $r$ and $\sigma$ are constant, and therefore the two possible changes in price are:
\begin{eqnarray*}
	\label{eq:S_t}
	d \ln S_{ut} = (r - \frac{1}{2}\sigma^2)dt + \sigma &  &
	d \ln S_{dt} = (r - \frac{1}{2}\sigma^2)dt - \sigma 
\end{eqnarray*}
The operator which simulates the changes in price over a given period, $\mathcal{C}$, consists of two registers. Register $\mathcal{B}$ prepares the Brownian motion variable $dW_t$ for each time step, while $\mathcal{P}$ holds the price of the underlying asset. The size of the latter determines how many binary decimal places will be used to store the final stock price distribution. Changes in price are calculated in log space using the Fourier basis until the end of this circuit. We use operator $\mathcal{D}$ inspired by \cite{draper2000addition} to add a constant to $\mathcal{P}$ in the Fourier basis. $\mathcal{C}$ consists of the following steps:

\begin{itemize}
	\item A single Hadamard gate to each qubit in $\mathcal{B}$ simulates Brownian motion.
	\item A QFT transforms $\mathcal{P}$ into the Fourier basis.
	\item $\mathcal{D}_{S_0}$ adds the log of the initial stock value $S_0$
	\item An application of $\mathcal{D}_{\ln S_{ut}}$ and $\mathcal{D}_{\ln S_{dt}}$ controlled on the $\ket{0}$ and $\ket{1}$ states respectively of each qubit in $\mathcal{B}$. See \ref{fig:ChangeCirc}.
	\item $\mathcal{D}_1$ converts $\mathcal{P}$ from log space to normal space using the first order Taylor approximation $x  \approx 1 + \ln x$.
	\item An iQFT transforms $\mathcal{P}$ back to the computational basis.
\end{itemize}

%\begin{lstlisting}[language=Python, breaklines=true, caption={Example \textit{Qiskit} code to generate the $\mathcal{D}$ gate for a given register P and input value.}, label=lst:dcode]
%qc = QuantumCircuit(P.size)	
%for i in range(P.size):
%	lam = value * (2**(P.size-2) * np.pi / (2**(i)))
%	qc.p(lam, i)
%qc.to_gate()
%\end{lstlisting}

Once the probability of all possible stock prices are loaded into $\mathcal{P}$, a quantum linear amplitude function $F \ket{x}$ computes 
\[ F\ket{x}\ket{0} = \sqrt{1 - \hat{f}(x)}\ket{x}\ket{0} + \sqrt{\hat{f}(x)}\ket{x}\ket{1}.
\]
where $\hat{f}$ is an affine mapping of the payoff of a European call option
\[f = \left\{ \begin{array}{lr}
	0, & \text{if } x \leq \text{strike}\\
	x - \text{strike}, & \text{if } x > \text{strike}
\end{array}\right\}. \]

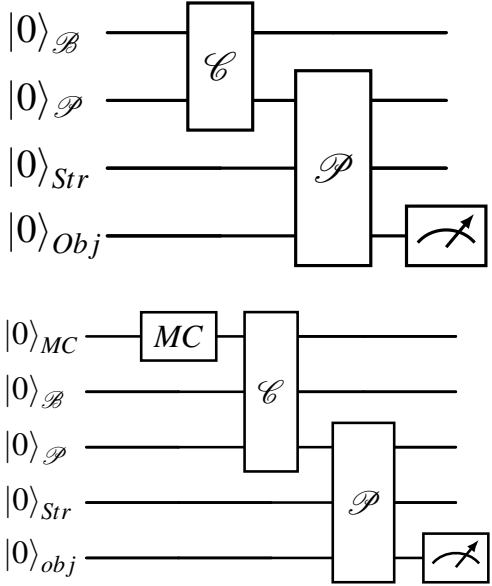
\begin{figure}[H]
	\centering
	\begin{tikzpicture}
		\node[scale=.8](circ1){
			\begin{adjustbox}{width=\columnwidth}
			\begin{quantikz}[row sep={0.6cm,between origins}, column sep={1.0cm,between origins}]
				\lstick[label style={text width=0.75cm, align=left}]{$\ket{0}_{\mathcal{B}}$} & \gate[2]{\mathcal{C}} & \qw & \qw \\
				\lstick[label style={text width=0.75cm, align=left}]{\ket{0}$_\mathcal{P}$} & & \gate[3]{\mathcal{P}} & \qw \\
				\lstick[label style={text width=0.75cm, align=left}]{$\ket{0}_{Str}$} & \qw & & \qw \\
				\lstick[label style={text width=0.75cm, align=left}]{$\ket{0}_{Obj}$} & \qw & &\meter{}
			\end{quantikz}
			\end{adjustbox}
		};
		\node [scale=.8] [below=0.1cm of circ1] {
			\begin{adjustbox}{width=\columnwidth}
			\begin{quantikz}[row sep={0.6cm,between origins}, column sep={1.0cm,between origins}]
				\lstick[label style={text width=0.75cm, align=left}]{$\ket{0}_{MC}$} & \gate{MC} & \gate[3]{\mathcal{C}} & \qw & \qw \\
				\lstick[label style={text width=0.75cm, align=left}]{$\ket{0}_{\mathcal{B}}$} & \qw & & \qw & \qw \\
				\lstick[label style={text width=0.75cm, align=left}]{$\ket{0}_{\mathcal{P}}$} & \qw & & \gate[3]{\mathcal{P}} & \qw \\
				\lstick{$\ket{0}_{Str}$} & \qw & \qw & & \qw \\
				\lstick[label style={text width=0.75cm, align=left}]{$\ket{0}_{obj}$} & \qw & \qw & &\meter{}
			\end{quantikz}
			\end{adjustbox}
		};
	\end{tikzpicture}
	\caption{\label{fig:DerivCirc} High level diagram of the Derivative Pricing circuit for the basic model (top) and regime switching (bottom). One time step of gate $\mathcal{C}$ for each model can be seen in Figure \ref{fig:ChangeCirc}. The payoff function $\mathcal{P}$ is explored in detail in \cite{woerner2019quantum}. It is worth noting that the first qubit of the MC register is not used in $\mathcal{C}$.}
\end{figure}

Following \cite{woerner2019quantum}, the mapping $\hat{f}$ is scaled down to leverage the small angle approximation $\sin^2\left(y + \frac{\pi}{4}\right) \approx y + \frac{1}{2}$ for sufficiently small $y$. The operator $F$ uses an additional register which is prepared to equal the strike price. Next, a quantum integer comparator is applied to $\mathcal{P}$ and the strike price register. The result of this comparator is stored in an ancilla qubit which is then used to control a Pauli polynomial rotation that applies $F\ket{x}$ to the objective qubit for all $\ket{x}$ greater than the strike price. Figure \ref{fig:DerivCirc} shows an overview of the complete derivative pricing circuit.

The Markov chain version of the model can be implemented via a similar circuit with the addition of the operator outlined in section \ref{sec:Markov} and a modification to $\mathcal{C}$. In the Markov chain version, the fourth step of $\mathcal{C}$ is converted from two controlled applications of $\mathcal{D}_{\ln S}$, to four multi-controlled applications of the same gate: one for each combination of the economic state variable $X_t$ and Brownian motion variable $dW_t$. See a comparison of the two gates in Figure \ref{fig:ChangeCirc}. 

\begin{table*}
	\begin{center}
		\begin{tabular}{lcccc}
			Regime Switching & $GE \sigma$ & $BE \sigma$ & $GE r$ & $BE r$ \\
			& & & &  \\
			No  & 0.2 + min(max(1.2t,0),0.1) & -- & 0.1 & -- \\
			Yes & 0.2 + min(max(1.2t,0),0.1) & 0.3 + min(max(1.2t,0),0.1) & 0.2  & 0.1 \\
		\end{tabular}
		\caption{\label{tab:DerivParams} Derivative pricing model parameters we use for the experiments whose results are reported in Table \ref{tab:CredRiskRes} with and without regime switching. We label the state dependent default probability risk free interest rate $r$ and volatility $\sigma$ for the good economy (GE) and bad economy (BE) regimes respectively. The volatility increases in the bad economy regime, while the risk free interest rate decreases. For the case without regime switching, we list variables in the GE column for simplicity.}
	\end{center}
\end{table*}
We test our algorithm using the parameters in Table \ref{tab:DerivParams}. All experiments used 6 time steps, $S_0$ = 1.0, T = six months, and strike prices 0.9, 0.95, 1.0, 1.05, and 1.1. We benchmark against an exact solution classically calculated with no sampling error, which was feasible thanks to the relatively short chain.  We show the MSE of our results for S = 128, B = 5, and N = 100 in Table \ref{tab:DerivMSE}.

\begin{table*}
	\begin{center}
		%	\vspace{1.0em}
		%		\footnotesize
		\begin{tabular}{lccc}
			Processor & No Switching & 1986-present & Synthetic  \\
			& & & \\
			Simulator & 2.1e-4 & 1.5e-4 & 3.4e-4 \\
			Quantum & 5.1e-3 & 3.8e-3 & 2.7e-3 \\
		\end{tabular}
		%--------------------------------------------------
		%	\noindent{\scriptsize The entries are the Mean-Squared Errors}
		\caption{\label{tab:DerivMSE} The MSE of the Derivative pricing models, once again run with and without regime switching on the noiseless \textit{ibmq\_simulator\_mps}, as well as the \textit{ibm\_sherbrooke} Eagle quantum processor with and without PEC. The row number in the Parameters column is in reference to Table \ref{tab:DerivParams}.}
	\end{center}
\end{table*}

%\section{Discussion}

Of the circuits outlined in this paper, the credit risk application is a stronger candidate for usable Amplitude Estimation result on NISQ processors, as demonstrated for the non-regime switching model in \cite{woerner2019quantum}. The payoff function for derivative pricing yields a measurement probability $prob$, which is converted into a value via the classical post processing
\[ \text{Price} = \frac{2 f_{max}}{\pi c} \left(prob - \frac{1}{2} + \frac{c \pi}{4} \right), \]
where $f_{max}$ is the highest payout possible given the strike price and the confines of the $\mathcal{P}$ register, and $c$ is a re-scaling constant required to make use of the small angle approximation. Choosing a smaller value for $c$ decreases the approximation error, however it increases the sampling error from QAE outlined in section \ref{sec:credit_dynam} since for a given change in measurement probability $\Delta prob$, the change in derivative price $\Delta \text{price} \propto 1/c$. On an ideal quantum processor, the optimal choice of $c$ results in a total error that scales with $\mathcal{O}\left(M^{-\frac{2}{3}}\right)$ as opposed to $\mathcal{O}\left(M^{-1}\right)$ for the credit risk example. More research needs to be done to determine the optimal choice of scaling constant when error from real quantum hardware is taken into account.

\section{A Practical Implementation Example \label{sec:credit_dynam}}
We consider a practical implementation of credit risk, computing the V@R for a portfolio of risky corporate bonds. More specifically, we look at a portfolio mix of Aaa and Baa bonds. We use indices to reflect the prices of these type of bonds and collect Aaa and Baa monthly series from 1986 until end of 2022 called respectively  $Aaa_t$ and $Baa_t$.\footnote{Data sources are for the Aaa series, \url{https://fred.stlouisfed.org/series/AAA10Y}, and for the Baa: \url{https://fred.stlouisfed.org/series/BAA10Y}} Again using the NBER business cycle chronology, we define $r^b_t$ = 1 + $Baa_t$ and  $r^a_t$ = 1 + $Aaa_t$ and compute means and standard deviations of $r^b_t$ and $r^a_t$ during NBER expansions only. We call the NBER expansions the good economy $GE$ states and recessions the $BE$ ones. And therefore we use this $me_a^G,$ $me^G_b,$ for the means of $a$ and $b,$ and $sd^G_a$ and finally  $sd^G_b$ for the standard deviations. We compute the same statistics for NBER recessions, call these $me^B_a,$ $me^B_b,$  $sd^B_a$ and  $sd^B_b.$ Finally, we compute the correlation between $r^b_t$ and $r^a_t$ during NBER expansions only, call it $\rho^G$ amd during recessions only $\rho^B.$ Define:
\begin{eqnarray*}
	\Sigma^G & = & \left[\begin{array}{cc}
		(sd^G_a)^2 & sd^G_a sd^G_b \rho^G \\
		sd^G_a sd^G_b \rho^G & 	(sd^G_b)^2
	\end{array}\right] \\
	\Sigma^B & = & \left[\begin{array}{cc}
		(sd^B_a)^2 & sd^B_a sd^B_b \rho^B \\
		sd^B_a sd^B_b \rho^B & 	(sd^B_b)^2
	\end{array}\right] \\
	\mu^G & = & \left[\begin{array}{c}
		me^G_a \\
		me^G_b
	\end{array}\right] \quad 
	\mu^B \,  = \, \left[\begin{array}{c}
		me^B_a \\
		me^B_b
	\end{array}\right] \\
\end{eqnarray*}
We use the generic stationary two-state Markov setting where the states are observable with Good and Bad outcomes and their associated transition probabilities.\footnote{Making the NBER chronology an observable state in real-time is a simplification. Usually the NBER calls recessions and expansion states with some time delay.}
Then a Markowitz mean-variance portfolio, see \cite{markowitz1959portfolio}, can be formulated as follows:
\begin{eqnarray*}
	\underset{w_{t+1}(X_t)}{\max} \left[ w_{t+1}(X_t)^\top \mu^G - \lambda  w_{t+1}(X_t)^\top \Sigma^G  w_{t+1}(X_t) \right] \times \\
	p(X_{t+1} = G|X_t)  \\
+	\left[ w_{t+1}(X_t)^\top \mu^B - \lambda  w_{t+1}(X_t)^\top \Sigma^B  w_{t+1}(X_t) \right] \times \\
	p(X_{t+1} = B|X_t) 	\\
	\text{subject to } \iota^\top w_{t+1}(X_t) = 1 
\end{eqnarray*}
where $\iota$ is a $n \times 1$ vector of ones ($n$ the number of assets), which has the following closed-form solution:
{\small
\begin{eqnarray*}
	w_{t+1}(X_t) & = & \frac{1}{2 \lambda} \left[(\Sigma^G)^{-1}(\mu^G + \nu^G\iota)p(X_{t+1} = G|X_t)\right] \\
	& & + \frac{1}{2 \lambda} \left[ (\Sigma^B)^{-1}(\mu^B + \nu^B\iota) p(X_{t+1} = B|X_t) \right] \\
	\nu^G & = & \frac{2 \lambda - \iota^\top (\Sigma^G)^{-1} \mu^G}{\iota^\top (\Sigma^G)^{-1} \iota} \\
	\nu^B  & =  & \frac{2 \lambda - \iota^\top (\Sigma^B)^{-1} \mu^B}{\iota^\top (\Sigma^B)^{-1} \iota} 
\end{eqnarray*}}
where we provide in Table \ref{NBERportfolioparms} a summary of the sample-based numerical values used in our computations. We are dealing with a stationary Markov chain with $p_{GB}$ = $p(X_{t+1} = B|X_t = G),$  $p_{BG}$ = $p(X_{t+1} = G|X_t = B).$ Therefore the portfolio weights take a stationary solution only depending on the state with for $w^G_a$ for Aaa bonds,  $w^G_b$ = 1 - $w^G_a$ for Baa bonds for the $G$ state , whereas for $B$ states we have $w^B_a$ and $w^B_b$ = 1 -  $w^B_a.$

\begin{table*}
	\begin{center}
		\begin{tabular}{lrclrclrclr }
		$me^G_a$ & 2.31 & & $me^G_b$ & 3.23 & & $me^B_a$ & 2.78 & & $me^G_B$ & 4.33 \\
		$sd^G_a$ & 0.42 & & $sd^G_b$ & 0.56 & & $sd^B_a$ & 0.62 & & $sd^B_b$ & 1.24 \\
		$\rho^G$ & 0.91 & & $\rho^B$ & 0.85 & & $p_{GB}$ & 0.0097 & & $p_{BG}$ & 0.11 \\
		$w^G_a$  & 0.81 & & $w^G_b$  & 0.19 & & $w^B_a$  & 1.3 & & $w^B_b$  & -0.3 \\
		\end{tabular}
		\caption{\label{NBERportfolioparms} Using the National Bureau of Economic Research US business cycle data from 1986 to 2020, the table contains the parameters used for a Markowitz mean-variance portfolio mix of Aaa and Baa bonds. Denote indices to reflecting the prices of these type of bonds as $Aaa_t$ and $Baa_t$ and define $r^b_t$ = 1 + $Baa_t$ and  $r^a_t$ = 1 + $Aaa_t.$ We call the NBER expansions the 'Good' states and recessions the 'Bad' ones. And therefore we use this $me_a^G,$ $me^G_b,$ for the mean returns of $a$ and $b$ bonds, and $sd^G_a$ and finally  $sd^G_b$ for the standard deviations. We compute the same statistics for NBER recessions, call these $me^B_a,$ $me^B_b,$  $sd^B_a$ and  $sd^B_b.$ We compute the correlation between $r^b_t$ and $r^a_t$ during NBER expansions only, call it $\rho^G$ amd during recessions only $\rho^B.$ Finally, $p_{GB}$ = $p(X_{t+1} = B|X_t = G),$  $p_{BG}$ = $p(X_{t+1} = G|X_t = B)$ and portfolio weights for $G$ are $w^G_a$ for Aaa bonds,  $w^G_b$ = 1 - $w^G_a$ for Baa bonds, whereas for $B$ $w^B_a$ and $w^B_b$ = 1 -  $w^B_a.$}
	\end{center}
\end{table*}

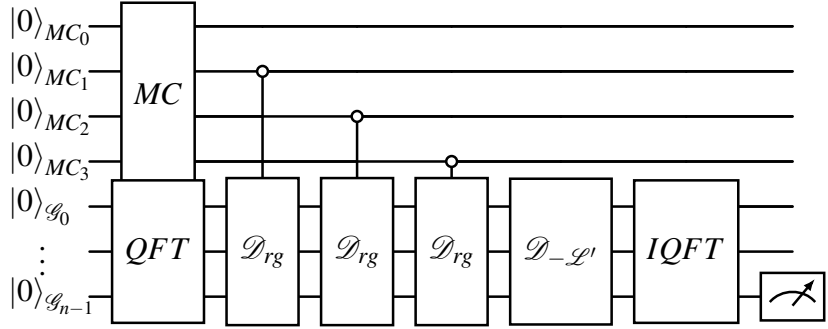
\begin{figure*}
	\centering
	\begin{tikzpicture}
		\node[scale=0.8]{
			\begin{adjustbox}{width=0.8\linewidth}
			\begin{quantikz}[column sep={0.25cm}, row sep={0.5cm,between origins}]
				\lstick[label style={text width=0.75cm, align=left}]{$\ket{0}_{MC_0}$} & \gate[4]{MC} & \qw & \qw & \qw & \qw & \qw & \qw  \\
				\lstick[label style={text width=0.75cm, align=left}]{$\ket{0}_{MC_1}$} &  & \octrl{3} & \qw & \qw & \qw & \qw & \qw \\
				\lstick[label style={text width=0.75cm, align=left}]{$\ket{0}_{MC_2}$} &  & \qw & \octrl{2} & \qw & \qw & \qw & \qw \\
				\lstick[label style={text width=0.75cm, align=left}]{$\ket{0}_{MC_3}$} &  & \qw & \qw & \octrl{1} & \qw & \qw & \qw \\
				\lstick[label style={text width=0.75cm, align=left}]{$\ket{0}_{\mathcal{G}_0}$} & \gate[wires=3, nwires=2]{QFT} & \gate[3, nwires=2]{\mathcal{D}_{rg}} & \gate[3, nwires=2]{\mathcal{D}_{rg}} & \gate[3, nwires=2]{\mathcal{D}_{rg}} & \gate[3, nwires=2]{\mathcal{D}_{-\mathcal{L}'}} & \gate[3, nwires=2]{IQFT} & \qw \\
				\lstick[label style={text width=0.75cm, align=center}]{\vdots}& & & & & & & \\
				\lstick[label style={text width=0.75cm, align=left}]{$\ket{0}_{\mathcal{G}_{n-1}}$} & & & & & & & \meter{} \\
			\end{quantikz}
			\end{adjustbox}
		};
	\end{tikzpicture}
	\label{fig:dynamCreditCirc}
	\caption{Diagram of a quantum circuit evaluating a three time step dynamic portfolio credit risk model. We reuse the $\mathcal{D}$ gate from the Derivative Pricing circuit. The number of qubits $n$ in $\mathcal{G}$ is equal to the number of binary digits needed to represent the largest possible growth with the desired fractional precision, plus one for the sign index.}
\end{figure*}

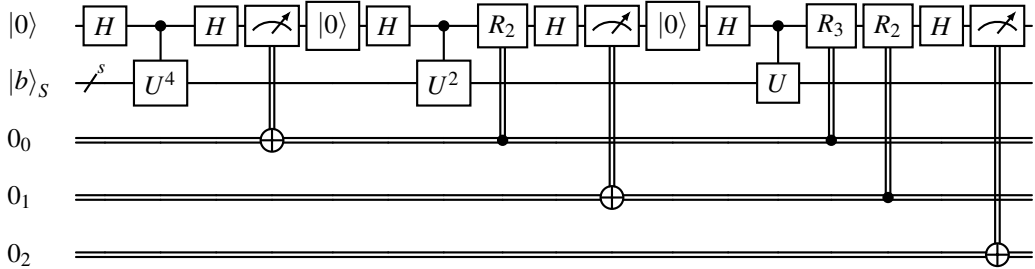
\begin{figure*}
	\centering
	\begin{tikzpicture}
		\node[scale=1]{
			\begin{adjustbox}{width=0.8\linewidth}
				\begin{quantikz}[column sep={0.1cm}, row sep={0.75cm,between origins}, wire types = {q,q,c,c,c}]
					\lstick[label style={text width=0.75cm, align=left}]{$\ket{0}$} & \gate{H} & \ctrl{1} & \gate{H} & \meter{} \wire[d][2]{c} &
					\gate{\ket{0}}  & \gate{H} & \ctrl{1} & \gate{R_2} & \gate{H} & \meter{} \wire[d][3]{c} & \gate{\ket{0}} & \gate{H} & \ctrl{1} & \gate{R_3} & \gate{R_2} & \gate{H} & \meter{} \wire[d][4]{c} & \\
					\lstick[label style={text width=0.75cm, align=left}]{$\ket{b}_{S}$} & \qwbundle{s} & \gate{U^4} &  &  &  &  & \gate{U^2} &  & & &  &  & \gate{U} & & & & & \\	
					\lstick[label style={text width=0.75cm, align=left}]{$0_0$} & &  &  & \targ{} & &  &  & \ctrl[vertical wire=c]{-2} & & &  &  & & \ctrl[vertical wire=c]{-2} & & & & \\
					\lstick[label style={text width=0.75cm, align=left}]{$0_1$} & &  &  &  &  & &  &  & & \targ{} &  &  & & & \ctrl[vertical wire=c]{-3} & & & \\
					\lstick[label style={text width=0.75cm, align=left}]{$0_2$} & &  &  &  & &  &  &  & & &  &  & & & & & \targ{} & \\			
				\end{quantikz}	
			\end{adjustbox}		
		};
	\end{tikzpicture}
	\label{fig:dynamCreditCirc}
	\caption{Curcuit diagram of QCL-QPE circuit with $n=3$ classical bits used to generate eigenvalue estimates. The classical bits take the place of the clock register of quantum bits in the canonical QPE circuit. This means that increasing the decimal bit precision of the estimates does not require additional qubits nor additional SWAP gates on limited connectivity hardware.}
\end{figure*}  

As with the static portfolio example, the problem of interest is to find $\mathcal{L}$ for a given $\alpha$, where $\alpha$ is the confidence interval that the loss over a given time horizon will not exceed $\mathcal{L}$. Before modeling the V@R of a mean variance of a dynamic portfolio managed with a given risk aversion, we perform a classical mapping to reduce the number of qubits needed for the quantum calculation. The expected growth of the portfolio over one month is $g_G = (w^G)^\top \mu^G$ in the good regime and $g_B = (w^B)^\top \mu^B$ in the BE regime. Let the relative growth be $rg = g_G - g_B$. We classically map $\mathcal{L}' = \mathcal{L} - \mathcal{L}_max$ where $\mathcal{L}_max = g_B \times T$ or the loss in the worst possible outcome within the confines of the model. The quantum portion of the algorithm computes $\alpha$ of $\mathcal{L}'$ for a portfolio that grows by $rg$ in the GE and does not change in the BE.  If $g_G < g_B$ for the portfolio being studied, then the place of GE and BE are swapped in the above mapping. 
Our circuit shown in figure \ref{fig:dynamCreditCirc} starts with a Markov Chain and a QFT of the total growth register $\mathcal{G}$. The number of qubits in $\mathcal{G}$ is equal to the number of fractional binary digits chosen plus $\log_2(rg*T) + 1$ rounded up to the next integer. This assures that $\mathcal{G}$ can accommodate the highest possible relative growth over the time period $T$ without overflow. Next, we use the GE state of each time step in $\mathcal{MC}$ to control an addition of $rg$ to $\mathcal{G}$ using the $\mathcal{D}$ operator outlined in section \ref{sec:deriv}. This creates a probability distribution of all possible growths. We then subtract $\mathcal{L}'$ from $\mathcal{G}$ and apply an IQFT to return to the computational basis. At the end of this circuit, the probability that $\mathcal{G}<0$ is equal to the alpha. Draper QFT addition uses Two's complement to store negative numbers, thus the sign qubit of $\mathcal{G}$ is the objective bit.  
The approximation error of repeatedly measuring the objective qubit is equivalent to that of a classical Monte Carlo. A theoretical quantum advantage can be achieved with Qantum Ampltidue Estimation (QAE). To apply the canonical QAE algorithm we map the problem of interest to a quantum operator, namely a unitary operator $\mathcal{A}$ acting on a register of $n$ + 1 qubits such that
\[
\mathcal{A} |0\rangle_{(n+1)} = \sqrt{1 - a} |\psi_0\rangle_n  |0 \rangle + \sqrt{a} | \psi_1 \rangle_n |1 \rangle
\]
for some normalized states  $| \psi_0 \rangle_n$ and $| \psi_1 \rangle_n,$ where $a$ $\in$ $[0,1]$ is unknown. QAE allows the efficient estimation of $a,$ i.e., the probability of measuring $|1\rangle$ in	the last qubit. This estimation is obtained with a Grover operator
\[
\mathbb{Q} = \mathcal{A} \mathbb{S}_0 \mathcal{A}^\dagger \mathbb{S}_{\psi_0},
\]
where $\mathbb{S}_0$ = 1 - 2 $|0\rangle \langle 0|$ and $\mathbb{S}_{\psi_0}$ =  1 - 2 $|\psi_0\rangle |0\rangle \langle \psi_0| \langle 0|,$
and which is a rotation of angle $2 \theta_a$ (where $a$ = $\sin^2(\theta_a)$) in the 2-dimensional space spanned by $|\psi_0\rangle_n  |0 \rangle$ and $| \psi_1 \rangle_n |1 \rangle.$ Recent work has established that QAE provides an estimation of $a$ with an estimation error which scales with 
$\mathcal{O}\left(M^{-1}\right)$ for large values of $M$ which is the number of quantum samples, equal to the number of applications of $\mathbb{Q}$. This is a quadratic speedup over the estimation error of a classical Monte Carlo method which scales with $\mathcal{O}\left(M^{-1/2}\right)$. The practical example circuit outlined above is prohibitively large to estimate with canonical QAE on current processors \footnote{Transpiling a canonical QAE circuit of the one year example for real backend requires more classical memory than is allotted to users of the IBM Quantum Lab \cite{IBM23}. We did not pursue further efforts to allocate memory to transpile said circuit, as this barrier along with the complexity of our state prep operator were sufficient evidence of the canonical circuit's infeasibility.}. To reduce the circuit size while still achieving a speedup, we follow the Iterative QAE (IQAE) algorithm outlined in \cite{Grinko_2021}. IQAE out performs the approximation error of canonical QAE, and its circuit demands less connectivity between qubits, which is particularly advantageous for our large state preparation circuit on superconducting processors. Iterative Amplitude estimation uses uncontrolled applications of $Q^k$, where the value of k is chosen after each iteration consisting of a given number of shots to maximize the Fisher information gained by each subsequent measurement. This process is repeated until sufficient information is collected make an estimate that the result is within $\epsilon$ of the $a$ with a set confidence level. We aimed for a $95\%$ confidence interval and chose to follow \cite{Grinko_2021} by conducting iterations of 100 shots. We chose $\epsilon = 0.1$ which exceeds the sampling error of cannonical QAE with 4 evaluation qubits. The simulated results of IQAE of our practical example are shown in Table \ref{tab:IQAE}.

 \begin{table*}
	\begin{center}
		%	\vspace{1.0em}
		%		\footnotesize
		\begin{tabular}{lcccccc}
			Time Steps & Loss & Ideal $\alpha$ & Estimate & CI & Ideal Complexity & Overhead\\
			& & & & & & \\
			3 & 2.3\% & 6.4\% & 7.7\% & 2.2\% & 1082 & 2.00 \\
			6 & 4.6\% & 4.5\% & 6.0\% & 1.9\% & 12932 & 1.88 \\
		\end{tabular}
		%--------------------------------------------------
		%	\noindent{\scriptsize The entries are the Mean-Squared Errors}
		\caption{\label{tab:IQAE} The estimates and confidence intervals (CI) from Iterative Amplitude Estimation of the practical dynamic portfolio calculation evaluated on the \textit{ibmq\_qasm\_simulator}. For each number of time steps, we estimate $\alpha$ for a chosen $\mathcal{L}$ whose true $\alpha$ is closest to 0.05. The ideal complexity is the number of two-qubit gates needed to run the circuit on a perfect connectivity processor which uses the same basis gates as \textit{ibmq\_sherbrooke}. The connectivity restraints of the real processor increase the complexity of the transpiled circuit by a factor of approximately 2, as shown in the overhead column.}
	\end{center}
\end{table*}

To evaluate the computational complexity of our models, we focus on the number of non-local gates required to implement the various forms of our practical example circuit because the error rate for single qubit gates is an order of magnitude less than that of two-qubit gates on the Eagle r3 processors we primarily used for our tests \cite{IBM23}. These processors unroll non-local interactions into a composition of Echoed Cross Resonance (ECR) gates and single qubit operations. The composition is optimized via an iterative heuristic process that eliminates redundancies outlined in \cite{runtime}. We study cannonical QAE with 4 evaluation qubits because that is where the approximation error surpasses a classical Monte Carlo with equal number of sampled. We only look at the iterative QAE circuit when k=1, which was the largest k needed to exceed the aforementioned approximation error. We determined that this was ideal for our proof of concept example to maintain an achievable circuit depth. While circuit depth constraints prevent us from utilizing the scaling advantage at this time, we are able to estimate a model with real industry applicability on a Quantum Processor in a manner that scales better than a classical Monte Carlo.

\section{Conclusion}

The purpose of this paper is to propose a versatile class of risk models based on a Markovian regime switching environment which are better suited for implementation of quantum computing algorithms to realistic financial market conditions. Obviously, we still have to cope with the limitations of NISQ hardware, and that includes error mitigation and constraints on the number of qubits. Nevertheless, the models discussed in this paper provide a road map for future developments in the area of Fintech applications of quantum computing.

\section{Acknowledgments}
The first author acknowledges the financial support from an IBM Global University Program Academic Award. We would like to thank Stefan Woerner for valuable comments and feedback on an earlier draft of the paper. We also thank Nik Stamatopoulos for insightful discussions.
\end{multicols}

\begin{multicols}{2}
%\setcounter{section}{0}
%\setcounter{equation}{0}
%\setcounter{table}{0}
%\setcounter{figure}{0}
%\renewcommand{\theequation}{A.\arabic{equation}}
%\renewcommand\thetable{A.\arabic{table}}
%\renewcommand\thefigure{A.\arabic{figure}}
%\renewcommand\thesection{A.\arabic{section}}

%\bigskip

%\begin{center}
%	{\Large\textbf{APPENDIX}}	
%\end{center}
%\bigskip

%\begin{center}
%	[TO BE INCLUDED]
%\end{center}

%\bigskip

\bibliography{References}
% \printbibliography{}

\end{multicols}

\end{document}